\documentclass[11pt]{article}
\newcommand{\Comment}[1]{{}}
\usepackage{amsfonts,amsthm,amsmath,amssymb,slashed}
\usepackage[textwidth = 430 pt, textheight = 630 pt]{geometry}

\Comment{\usepackage{color}
\definecolor{MyDarkBlue}{rgb}{0.15,0.15,0.45}
\usepackage[linktocpage=true]{hyperref}
\hypersetup{
colorlinks=true,
citecolor=MyDarkBlue,
linkcolor=MyDarkBlue,
urlcolor=MyDarkBlue,
}

\usepackage[numbers,sort&compress]{natbib}
\usepackage{hypernat}}
\usepackage{graphicx,dsfont}
\usepackage{cite}

\newcommand\ignore[1]{}
\def\one{{\,\hbox{1\kern-.8mm l}}}

\def\({\left(}
\def\){\right)}

\newcommand{\Cset}{{\,\,{{{^{_{\pmb{\mid}}}}\kern-.45em{\mathrm C}}}}}

\newcommand{\be}{\begin{equation}}
\newcommand{\bea}{\begin{eqnarray}}

\newcommand{\ee}{\end{equation}}
\newcommand{\eea}{\end{eqnarray}}

\begin{document}

\renewcommand{\thefootnote}{\fnsymbol{footnote}}

\makeatletter
\@addtoreset{equation}{section}
\makeatother
\renewcommand{\theequation}{\thesection.\arabic{equation}}

\rightline{}
\rightline{}
   \vspace{1.8truecm}

%\begin{flushright}
% preprint nrs.
%\end{flushright}

\vspace{10pt}

%%%%%%%%%%%%%%%%%

\begin{center}
{\LARGE \bf{\sc Conformal Symmetries of Adiabatic Modes in Cosmology}}
\end{center} 
 \vspace{1truecm}
\thispagestyle{empty} \centerline{
{\large \bf {\sc Kurt Hinterbichler${}^{a,}$}}\footnote{E-mail address: \Comment{\href{mailto:kurthi@physics.upenn.edu}}{\tt kurthi@physics.upenn.edu}},
{\large \bf {\sc Lam Hui${}^{b,}$}}\footnote{E-mail address: \Comment{\href{mailto:lhui@astro.columbia.edu}}{\tt lhui@astro.columbia.edu}}
{\bf{\sc and}} 
{\large \bf {\sc Justin Khoury${}^{a,}$}}\footnote{E-mail address: \Comment{\href{mailto:jkhoury@sas.upenn.edu}}{\tt jkhoury@sas.upenn.edu}}
                                                           }

\vspace{1cm}

\centerline{{\it ${}^a$ 
Center for Particle Cosmology, Department of Physics and Astronomy,}}
 \centerline{{\it University of Pennsylvania, Philadelphia, PA 19104, USA}}
 
\vspace{.5cm}
\centerline{{\it ${}^b$ 
Physics Department and Institute for Strings, Cosmology and Astroparticle Physics,}} \centerline{{\it 
Columbia University, New York, NY 10027, USA}}

\vspace{2truecm}

%%%%%%%%%%%%%%%%%
\thispagestyle{empty}

\centerline{\sc Abstract}

\vspace{.4truecm}

\begin{center}
\begin{minipage}[c]{380pt}
{\noindent We remark on the existence of non-linearly realized conformal symmetries for scalar adiabatic perturbations in cosmology. These 
conformal symmetries are present for any cosmological background, beyond any slow-roll or quasi-de Sitter approximation. The dilatation transformation shifts
the curvature perturbation by a constant, and corresponds to the well-known symmetry under spatial rescaling. We argue that the scalar sector is also invariant
under special conformal transformations, which shift the curvature perturbation by a term linear in the spatial coordinates. We discuss whether these conformal symmetries
can be extended to include tensor perturbations. Tensor modes introduce their own set of non-linearly realized symmetries. We identify an infinite set of large
gauge transformations which maintain the transverse, traceless gauge condition, while shifting the tensor mode non-trivially.}
\end{minipage}
\end{center}

\vspace{.5cm}

\setcounter{page}{0}
\setcounter{tocdepth}{2}

\newpage

%\tableofcontents
\renewcommand{\thefootnote}{\arabic{footnote}}
\setcounter{footnote}{0}

\linespread{1.1}
\parskip 4pt

%{}~
%{}~

\section{Introduction}
\label{intro}
\ \ \ \ \ 

Inflation~\cite{Starobinsky:1979ty,Guth:1980zm,Albrecht:1982wi,Linde:1981mu} proposes that the universe underwent a short phase of approximate de Sitter expansion shortly after the big bang.
The space-time is not exactly de Sitter, since inflation must eventually end. But for spectator fields fluctuating on this background, {\it i.e.} fields with negligible backreaction, de Sitter space is an excellent approximation. 

It is well known that de Sitter space enjoys 10 isometries, whose generators form the $so(4,1)$ algebra. At late times, these de Sitter isometries act as conformal transformations on the boundary $\mathbb{R}^3$ at infinity. In addition to spatial translations and rotations, the conformal algebra on $\mathbb{R}^3$ includes the dilatation and 3 special conformal transformations (SCTs), which infinitesimally in the parameters $\lambda$ and $b^i$ read
\bea
\nonumber
x^i &\rightarrow & (1 + \lambda) x^i \,, \\
x^i & \rightarrow & x^i + 2\, \vec x\cdot \vec b\, x^i -b^i \vec{x}^2\,.
\label{confsym}
\eea
This forms the basis of the proposed dS/CFT correspondence~\cite{Strominger:2001pn}. In particular, since correlation functions of spectator fields
are evaluated at late times, when the relevant modes are well outside the horizon, they should reflect these conformal
symmetries~\cite{Antoniadis:1996dj,Antoniadis:2011ib}. The scale invariance of the two-point function of nearly massless fields is a consequence of
dilatation invariance. Higher-point functions are similarly constrained by conformal invariance. This fact has been used recently to
derive the form of correlation functions for gravitons~\cite{Maldacena:2011nz} and spectator scalar fields~\cite{Creminelli:2011mw}.

Inflaton perturbations are a key exception. Because the inflaton is the agent that governs when inflation ends, perturbations along the inflaton
trajectory are sensitive to departures from de Sitter space. The inflaton picks out a preferred time slicing of de Sitter, which spontaneously breaks the $so(4,1)$ global symmetries.
As a result, inflaton correlation functions cannot be invariant under the full $so(4,1)$ symmetries.

In this note, we show that, for certain quantities, a conformal symmetry is (at least formally) at play in any cosmological Friedmann-Robertson-Walker (FRW) system with a single field, not restricted to quasi-de Sitter inflation.  By ``single field" we mean one degree of freedom responsible for driving the background and for generating density perturbations (this includes hybrid inflationary models~\cite{Linde:1993cn}).  If these symmetries are true physical symmetries of the theory, then scalar perturbations in single-field cosmology can be thought of as describing the spontaneous symmetry breaking pattern
\be
so(4,1)\rightarrow {\rm spatial} \; {\rm rotations} + {\rm translations}\,.
\label{symbreak}
\ee

This is most transparent in comoving gauge, where the field is unperturbed ($\phi = \phi(t)$) and the spatial metric is conformally flat
\be\label{ansatz1}
h_{ij} = a^2(t) e^{2\zeta(\vec{x},t)}\delta_{ij}\,. 
\ee
(We restrict our attention to scalar perturbations for now; note that
$\zeta$ is not necessarily small.) In this gauge, scalar perturbations are fully encoded in the curvature perturbation $\zeta$~\cite{Bardeen:1983qw,Salopek:1990jq}, which as we will see non-linearly realizes the spatial dilatation and special conformal symmetries~(\ref{confsym}), with infinitesimal transformations:
\bea
\nonumber
\delta_\lambda \zeta &=& \lambda (1+ \vec{x}\cdot \vec{\nabla}\zeta) \;; \\
\delta_{\vec{b}}\zeta  &=& 2\vec{b}\cdot \vec{x} + \left(2\vec{b}\cdot \vec{x}x^i -\vec{x}^2b^i \right)\partial_i\zeta\,. 
\label{delzeta}
\eea
This is essentially because the expression \eqref{ansatz1} is in the form of a metric conformal to flat space, so conformal transformations will preserve this form.  This does not contradict the usual notion that ``comoving gauge completely fixes the gauge," since this assumes that gauge transformations fall off at infinity --- conformal transformations clearly do not.  As such, the conformal transformations~(\ref{delzeta}) are residual diffeomorphisms mapping field configurations which fall off at infinity into those which do not. However, we will show in Sec.~\ref{adiamodes} that they can be extended to transformations which do fall off at infinity, and hence generate new physical solutions (adiabatic modes). 

The curvature perturbation can be recognized as the Goldstone boson ({\it i.e.}, the dilaton) for the symmetry breaking pattern~(\ref{symbreak})\footnote{It is well-known
that the usual counting of Goldstone bosons fails for space-time symmetries~\cite{Low:2001bw}. Although~(\ref{symbreak}) describes 4 broken
symmetries (dilatation + 3 SCTs), there is only one Goldstone mode.}. Remarkably, the non-linear realization of $so(4,1)$ does not assume slow roll,
and in fact applies to {\it any} cosmological background, not just de Sitter. The symmetry breaking pattern~(\ref{symbreak}) of interest
applies generally to adiabatic perturbations on any background, such as the late-time growth of density perturbations in a matter-dominated universe.

It is well known that even when symmetries of a theory are spontaneously broken and global conserved charges do not exist, local currents still exist, and can have important implications for correlation functions.  Spontaneously broken global symmetries can
lead to Ward identities, relating $N$ to $N-1$ point correlation functions. The classic example are the soft-pion theorems of chiral perturbation theory~\cite{softpion}. 

Correspondingly, if the symmetries we discuss here are truly physical symmetries --- we
will discuss the physical nature of these symmetries which
fundamentally have a
gauge origin in Sec.~\ref{discussion} ---
there will be ``soft-$\zeta$" theorems for inflationary correlation functions.  (Ward identities for conformal symmetries generically involve quantum anomalies~\cite{Coleman:1970je},
but our focus is on tree-level relations.)
The consistency relation~\cite{Maldacena:2002vr,Creminelli:2004yq,Cheung:2007sv} may already provide a known example of such theorems.
This relates the 3-point function in the squeezed limit ($k_1 \ll k_{2},k_3$) to the 2-point function:
\begin{equation}
\lim_{\vec{k}_1 \rightarrow 0} \langle \zeta_{\vec{k}_1} \zeta_{\vec{k}_2}\zeta_{\vec{k}_3}\rangle = -(2\pi)^3\delta^3(\vec{k}_1 + \vec{k}_2 + \vec{k}_3) (n_s - 1)P_{k_1}P_{k_3}\,.
\label{consis}
\end{equation}
This is a powerful test of single field inflation --- observing a violation of~(\ref{consis}) (through detection of a significant primordial $f_{\rm NL}^{\rm local}$) would
immediately rule out {\it all} single field inflationary models, and point to multi-field dynamics. The consistency relation has been established through
explicit calculations~\cite{Maldacena:2002vr,Cheung:2007sv} and can be understood using ``background-wave" arguments~\cite{Maldacena:2002vr,Creminelli:2004yq}.

In a forthcoming paper, we will show that the consistency relation derives from the Ward identity associated with the dilatation symmetry. Indeed, the
squeezed limit, with one external $\zeta$ taken to have very low momentum, is strikingly reminiscent of the soft-pion limit. Similarly, the Ward identity associated
with the broken SCTs will result in a novel consistency relation. 
This relation has been derived using the background-wave method
in a recent paper by Creminelli, Nore\~na and Simonovi\'c ~\cite{Creminelli:2012ed} (who have kindly shared
with us their preprint ahead of submission). There is also a parallel investigation by Goldberger, Hui and Nicolis
using the operator product expansion and path integral.

It is natural to ask whether the conformal symmetries can be generalized to include tensor perturbations, in such a way that the transverse, traceless (TT) gauge
condition is preserved. We will see that this can be done for the dilatation symmetry, to all orders in the tensor perturbations. Special conformal transformations,
on the other hand, result in a departure from the TT conditions, and hence are a good symmetry of the scalar sector only. Tensor perturbations, meanwhile, introduce
their own set of non-linearly realized symmetries. At linear order in the tensors but non-perturbatively in the scalars, we will find an infinite number of large
gauge transformations that shift the tensor mode while preserving the
TT gauge condition. (In this paper, by {\it large} gauge
transformations, we mean transformations that do not vanish at
infinity.) The simplest such transformation is an anisotropic scaling of coordinates,
corresponding to a shift symmetry for the tensors, but there will be an infinite set of $x$-dependent transformations as well.

\section{Set up}
\label{setup}

The starting point is the action describing general relativity (GR) plus a general, minimally coupled single-derivative scalar field,
\be
S=\int {\rm d}^4x \sqrt{-g}\left[{M_{\rm Pl}^2\over 2} R + P(X,\phi) \right]\,,
\label{actionstart}
\ee
where $X = -(\partial\phi)^2/2$. For a canonical scalar field with potential, for instance, $P(X,\phi) = X - V(\phi)$.
For a generic choice of $P(X,\phi)$, this action has no global symmetries --- there are only the gauge symmetries of
diffeomorphism invariance.  The change in the fields by an infinitesimal spacetime diffeomorphism, generated by a vector field $\xi^\mu(x)$, is 
\bea 
\delta g_{\mu\nu}=\xi^\rho\partial_\rho g_{\mu\nu}+\partial_\mu\xi^\rho\, g_{\rho\nu}+\partial_\nu\xi^\rho\, g_{\mu\rho}, \ \ \  \delta \phi= \xi^\mu\partial_\mu\phi\,.
\eea

The derivation of inflationary perturbation theory proceeds by going to ADM variables $N,N^i,h_{ij}$~\cite{Maldacena:2002vr},
\be
g_{\mu\nu}= \left(\begin{array}{c|c}- N^2+N^i N_{i}  & N_{i} \\ \hline N_{i} & h_{ij}  \end{array}\right)\,,
\ee 
where spatial indices are raised and lowered with the induced spatial metric $h_{ij}$. In these variables, the action~(\ref{actionstart}) is
\be 
S = \int {\rm d}^4x N\sqrt{h}\left[\frac{M_{\rm Pl}^2}{2}\(R^{(3)}+K_{ij}K^{ij}-K^2\)+P\left({1\over 2N^2}\(\dot \phi-N^i\partial_i\phi\)^2-\frac{1}{2}\(\nabla\phi\)^2,\phi\right)\right]\,,
\label{admaction}
\ee
where
\be
K_{ij}={1\over 2N}\(\dot h_{ij}-D_iN_j-D_jN_i\) 
\ee
is the extrinsic curvature tensor of constant-time surfaces, and the spatial covariant derivative $D_i$ is that of $h_{ij}$. 

Splitting the gauge parameter $\xi^\mu$ into a time component and space components,
\be 
\xi^\mu=\left(\xi,\xi^i\right)\,,
\ee
the gauge transformations of the ADM variables under which~(\ref{admaction}) is invariant are
\bea 
\delta h_{ij}&=&D_i\xi_j+D_j\xi_i+\xi\dot{h}_{ij}+N_i \partial_j\xi+N_j \partial_i\xi\,,\label{metrictrans}\\
\delta N^i&=&\xi^j\partial_j N^i-\partial_j \xi^i\ N^j+{{\rm d}\over {\rm d}t}\left(\xi N^i\right)+\dot{\xi}^i-\left( N^2h^{ij}+N^i N^j\right)\partial_j\xi \label{shifttrans}\, ,\\
\delta N&=&\xi^i\partial_i N+{{\rm d}\over {\rm d}t}\left(\xi N\right)-NN^i\partial_i\xi\,.
\label{lapsetrans}
\eea

With respect to a background solution $a(t)$, $\phi(t)$ satisfying the homogenous equations of motion (assuming spatial flatness),
\be 
3M_{\rm Pl}^2H^2= 2P_{,X} X - P \,; \ \ \ \  \frac{{\rm d}}{{\rm d}t} \left(a^3P_{,X} \dot{\phi}\right) = a^3P_{,\phi} \,,
\label{scalarfriedman}  
\ee
with $X = \dot{\phi}^2/2$, the co-moving (or uniform-density) gauge choice is
\be 
\phi=\phi(t)\,;\ \ \ \ h_{ij}=a^2(t)e^{2\zeta(\vec{x},t)}\(e^ \gamma\)_{ij}\,;\ \ \ \ \gamma^i_{\ i}=0\,;\ \ \ \ \partial_i\gamma^i_{\ j}=0\,,
\label{comovgauge}
\ee
with indices on the transverse traceless metric perturbation
$\gamma_{ij}$ raised and lowered with $\delta_{ij}$. This gauge choice was used by Maldacena in his classic paper~\cite{Maldacena:2002vr},
and is closely related to the conformal decomposition introduced by York~\cite{York:1971hw} and Lichnerowicz~\cite{lich}.
In this gauge, scalar inhomogeneities are captured by the curvature perturbation $\zeta$, while tensor modes are encoded in $\gamma_{ij}$.
If all the fields and gauge parameters are assumed to die off appropriately at spatial infinity, then this gauge choice completely fixes the gauge,
leaving no residual symmetries.

\section{Scalar symmetries}
\label{scalarsyms}

Even though the gauge choice~(\ref{comovgauge}) completely fixes the
gauge with respect to gauge parameters that fall off at spatial
infinity, there can still be room for gauge transformations involving
gauge parameters which do not die off at infinity. In this Section we
will see that conformal transformations on spatial slices, which
certainly do not fall off at spatial infinity, are formally residual
symmetries of the scalar sector. 
These generate non-trivial transformations
for $\zeta$, under which the scalar action must be invariant.
A word about our notation before we proceed: hereafter, unless otherwise stated,
we assume spatial indices are raised/lowered using $\delta_{ij}$ ---
this was implicitly assumed in Sec. \ref{intro} while the general
discussion in Sec. \ref{setup} used the full metric.

Suppose we are only interested in leading order correlators of $\zeta$, {\it i.e.}, those involving no graviton propagators in the diagrams.  Then we may set all graviton excitations to zero, $\gamma_{ij}\rightarrow 0$, and use the action~(\ref{admaction}) with \eqref{comovgauge} and
\be h_{ij}=a^2(t) e^{2\zeta(\vec{x},t)}\delta_{ij}\, ,
\label{ansatz}
\ee 
where $\zeta$ is not necessarily small.
There is a symmetry under which the action is
invariant while respecting the gauge choice in (\ref{ansatz})--- a
spatial conformal symmetry which acts non-linearly on $\zeta$.  

To see this, consider a spatial diffeomorphism, which may depend on time, $\xi^\mu=\(0,\xi^i(t,\vec x)\)$. 
Since this diffeomorphism is purely spatial, it does not change the gauge choice $\phi=\phi(t)$. 
Under the restriction~(\ref{ansatz}) to the scalar sector, if a variation $\delta\zeta$ can be chosen such that 
\be
\delta \left(a^2(t)e^{2\zeta(\vec{x},t)}\delta_{ij}\right)={\cal L}_{\vec\xi}\left(a^2(t)e^{2\zeta(\vec{x},t)}\delta_{ij}\right)\,,
\label{symchoicereq}
\ee
and if the lapse and shift transform as in~(\ref{shifttrans}) and~(\ref{lapsetrans}), then the action~(\ref{admaction}) with $h_{ij}=a^2(t) e^{2\zeta(\vec{x},t)}\delta_{ij}$ will be invariant, since the original action~(\ref{actionstart}) was diffeomorphism invariant. 
By the chain rule, this reduces to the requirement
\be
\delta\zeta \,\delta_{ij} = \xi^k\partial_k\zeta \,\delta_{ij} + \frac{1}{2}{\cal L}_{\vec\xi}\delta_{ij}\,.
\label{symchoicereq2}
\ee
Suppose $\xi^i$ is chosen to be a conformal Killing vector on $\mathbb{R}^3$, {\it i.e.},
\be  \label{CKVeq}
{\cal L}_{\vec{\xi}}\, \delta_{ij}=\partial_i\xi_j + \partial_j\xi_i = {2\over 3} \partial_k\xi^k\,\delta_{ij}\,.
\ee
Then~(\ref{symchoicereq2}) will be satisfied, and hence the
ansatz~(\ref{ansatz}) will satisfy the transformation law of a
diffeomorphism, if $\zeta$ is assigned the transformation law
\be
\delta \zeta=\xi^i\partial_i\zeta+{1\over 3}\partial_i\xi^i\,.
\label{zetatrans}
\ee
Therefore, under the transformation~(\ref{zetatrans}) for $\zeta$, together with the transformation rules~(\ref{shifttrans}) and~(\ref{lapsetrans}) for the shift vector and lapse function,
\be \delta N^i={\cal L}_{\vec\xi}N^i+\dot{\xi}^i,\ \ \  \delta N={\cal L}_{\vec\xi} N, \ee 
the GR action with the ansatz~(\ref{ansatz}) (which is nothing but the $\zeta$-action of~\cite{Maldacena:2002vr} before $N$ and $N^i$ are integrated out) is invariant for an arbitrary spatial conformal Killing vector $\xi^i$.  Note that the transformation of the lapse is linear, while the transformation of the shift contains a non-linear piece only if the transformation is time-dependent.

It is important to emphasize that no slow-roll or quasi-de Sitter approximations have been made. The conformal symmetries
identified above are completely general and hold in any FRW
background. Each symmetry will formally produce a local conserved current and a conserved charge via Noether's theorem.

There are 10 linearly independent spatial conformal killing vectors on $\mathbb{R}^3$.  Six of these are the rotations and translations, satisfying $\partial_i\xi^i=0$, for which the transformations on $\zeta$ are the usual linear rotations and translations. One of the vectors is the dilatation, $\xi^i(t,\vec x)=\lambda(t) x^i$, parametrized by $\lambda(t)$ (recall that we are allowing the transformations to be time-dependent), for which the transformation law reads
\be 
\delta_\lambda \zeta = \lambda(t) (1+ \vec{x}\cdot \vec{\nabla}\zeta)\,.
\label{zetadil}
\ee
The non-linear part of this transformation is a shift of $\zeta$.  Finally, there are the three spatial special conformal killing vectors, $\xi^i(t,\vec x)=2b^j(t)x_j x^i-\vec x^2b^i(t)$, parametrized by $b^i(t)$, for which the transformation law reads
\be 
\delta_{\vec{b}}\zeta  = 2\vec{b}(t)\cdot \vec{x} + \left(2\vec{b}(t)\cdot \vec{x}\, x^i -\vec{x}^2b^i(t) \right)\partial_i\zeta\,.
\label{zetaSCT}
\ee
The non-linear part of this transformation is a galileon shift \cite{Nicolis:2008in} of $\zeta$. 

We have checked explicitly that the action for $\zeta$ up to cubic order is invariant under
the non-linearly realized dilatation and SCT symmetries.  Once the lapse and shift are integrated out, the action is still local in time (but non-local in space), so there will continue to be a formal
conserved charge from Noether's theorem (though it may be infinite in the infinite volume limit), but not necessarily a {\it local}
conserved current in terms of $\zeta$ alone.

The current and charge are in fact easiest to write down without integrating
out $N$ and $N^i$. The charge density $j^0$ is particularly simple
for our symmetry transformations which
are purely spatial, and because the Lagrangian
does not depend on $\dot N$ and $\dot N^i$, we have
\be
j^0(x) =  \frac{\delta{\cal L}}{\delta \dot{\zeta}(x)} \delta\zeta(x) = \frac{1}{2} \{\Pi(x), \delta \zeta(x) \} \,,
\ee
where $\Pi \equiv \delta{\cal L}/\delta \dot{\zeta}$ is the canonical
momentum conjugate to $\zeta$ (and we have symmetrized the final term so that the resulting operator will be Hermitian in the quantum theory).
The integrated charge over an infinite volume is formally infinite, just as in theories with massless Goldstone bosons.

The conformal transformations identified above are so far pure gauge transformations. For the transformations~(\ref{zetadil}) and~(\ref{zetaSCT}) to have consequences on correlation functions involving physical modes,
the non-linear part of the transformations should be extendable to physical perturbations (adiabatic modes) with suitable fall-off
behavior at infinity~\cite{Weinberg:2008zzc,Weinberg:2003sw}.  In the next section we will argue 
that a subset of the transformations, namely time-independent dilatations and a diagonal combination of time-independent SCTs and time-dependent translations, can be extended to physical solutions.

\section{Adiabatic modes}
\label{adiamodes}

The symmetries we have identified are formally residual diffeomorphism symmetries of the action in co-moving gauge.  As is usual, we imagine restricting physical field configurations to satisfy some sort of asymptotic falloff conditions at spatial infinity.  For these configurations, co-moving gauge completely fixes the gauge.  The conformal transformations considered here do not fall off at spatial infinity, and therefore take field configurations which fall off at infinity into those which do not, so strictly speaking they are not symmetries.  However, some of them can be thought of as the $k\rightarrow 0$ limit of transformations which do fall off at infinity, and therefore generate new physical solutions.

Following an argument we have learned from
Creminelli, Nore\~na and Simonovi\'c
\cite{Creminelli:2012ed}, who generalized earlier work by Weinberg~\cite{Weinberg:2003sw,Weinberg:2008zzc},
it is possible to show that modes generated by the above symmetry
transformations
can, at least to linear order, be smoothly extended to finite momenta, that is, they can be
thought of as 
limits of field configurations which satisfy falloff conditions at 
asymptotic spatial infinity.  
This is an important check, since for example in the context
of consistency relations, these symmetry generated modes are
treated as the low momentum limit of physical perturbations.
Let us outline that argument here.

To be extendible to a physical mode, a configuration must satisfy the
equations of motion away from zero momenta, that is, it cannot ``accidentally'' solve the equations simply because they are being hit with spatial derivatives (which vanish as $k\rightarrow 0$).   The only equations for which this may happen are the constraint equations of GR, since the evolution equations have terms with second time derivatives and hence no spatial derivatives (since all the equations are second order).

The constraint equations are gauge invariant,
and hence are invariant under the transformations \eqref{metrictrans}, \eqref{shifttrans},
\eqref{lapsetrans}.  We will work to linear order, so we only need the non-linear part of the transformation laws (though it should be possible to extend the arguments to non-linear order, along the lines of \cite{Weinberg:2008nf}),
\be\label{lineartrans}
\delta \zeta={1\over 3}\partial_i\xi^i, \ \ \ \delta N^i=\dot{\xi}^i,\ \ \ \delta N=0.
\ee
Recall that the shift acquires a non-linear transformation when the conformal
Killing vector is time-dependent.  

First consider the momentum constraint. At first order in perturbation theory and in co-moving gauge, it reads
\be 
2\partial_i\left(H  N_1-\dot\zeta\right)-{1\over 2}\vec\nabla^2 N_i+{1\over 2}\partial_i\left(\partial_jN^j\right)=0\,.
\label{linearmometconst}
\ee
This is preserved by the transformations (\ref{lineartrans}), as can be seen by using the relation $\vec\nabla^2\xi_i=-{1\over 3}\partial_i\left(\partial_j\xi^j\right)$,
which follows by taking the divergence of the conformal Killing equation~\eqref{CKVeq}.  
Taking the divergence of \eqref{linearmometconst}, the terms involving the shift drop out, and we find
\be 
\vec \nabla^2\left(HN_1-\dot\zeta\right)=0\,.
\label{momcons}
\ee
The key point is that modes generated by the sort of
transformations we are interested in might 
satisfy momentum constraints like this
trivially, simply because they correspond to $k=0$.
To ensure that these modes are the $k \rightarrow 0$ limit of
physical modes, we must ensure $HN_1 - \dot \zeta = 0$ so
that the momentum constraints are satisfied even at a finite $k$.
For a mode of the form of the transformation (\ref{lineartrans}), we must have 
\be 
\frac{\partial}{\partial t} \left(\partial_i \xi^i\right)=0\,.
\ee
Thus the requirement that our mode be physical, in the sense that it
be extendible to a mode at finite $k$ with good boundary behavior,
forces the dilatation and special conformal transformations ({\it i.e.}, those with $\partial_i\xi^i\not=0$) to be time independent.

Meanwhile, the Hamiltonian constraint, at first order in perturbation theory, is
\be
\partial_i \left(\partial^i\zeta + HN^i\right) + \frac{a^2\dot{H}}{c_s^2}N_1 + 3a^2H\left(HN_1-\dot{\zeta} \right) = 0 \,,
\label{hamcons}
\ee
where $c_s^2 = P_{,X}/(P_{,X} + 2P_{,XX}X)$ is the sound speed of propagation.
At finite $k$,~(\ref{momcons}) implies that the last term vanishes. Focusing on the first term, $\partial_i\zeta$ transforms linearly under spatial dilatations,
and so does $N^i$, hence~(\ref{hamcons}) imposes no further constraint on dilatations. For (time-independent) spatial SCTs, on the other hand, $\partial_i\zeta$ transforms
non-linearly, whereas $N^i$ transforms non-linearly only if there is time dependence in the transformation. In order for the first term to be invariant at finite $k$,
the SCTs must be accompanied by a compensating time-dependent translation, $x^i \rightarrow x^i  + c^i(t)$, under which $\delta N^i = \dot{c}^i$ and $\delta\zeta = 0$.
Therefore only the diagonal combination,
\be
\xi^i(t,\vec x)=2\vec{b}\cdot\vec{x}\, x^i-\vec{x}^2b^i-2b^i\int^t{\rm d}t'{1\over H(t')}\,,
\ee
 of SCTs and time-dependent translations is extendible to finite $k$, and the other combinations are not.

\section{Tensor symmetries}
\label{tensyms}

Next we would like to understand whether the scalar symmetries identified above can be generalized to include tensors. We will see that dilatation invariance can be extended to all orders in tensor perturbations, whereas SCTs take us away from the transverse, traceless gauge. We will also find that tensors introduce new non-linearly
realized symmetries of their own.

\subsection{Tensor perturbations and conformal transformations}

The comoving gauge spatial metric including scalar and tensor perturbations is
\be
h_{ij}=a^2(t)e^{2\zeta(\vec{x},t)}\(e^ \gamma\)_{ij}\,,
\ee
where $\gamma_{ij}(\vec{x},t)$ is a transverse, traceless tensor, 
\be
\partial^i\gamma_{ij}=\gamma^i_{\ i}=0\,. 
\label{TT}
\ee
As emphasized in Sec. \ref{scalarsyms}, all spatial indices are raised and lowered with $\delta_{ij}$.

Let us first discuss the conformal symmetries at linear order in $\gamma_{ij}$. Generalizing~(\ref{symchoicereq}), we
seek a transformation rule such that 
\be
\delta \left(a^2(t)e^{2\zeta(\vec{x},t)}\left(\delta_{ij} + \gamma_{ij}\right) \right)={\cal L}_{\vec\xi}\left(a^2(t)e^{2\zeta(\vec{x},t)}\left(\delta_{ij} + \gamma_{ij}\right)\right)\,.
\label{symchoicereqtensors}
\ee
Moreover, to describe a symmetry for tensor modes, $\delta\gamma_{ij}$ should be transverse and traceless.
Substituting~(\ref{symchoicereq}) and the $\zeta$ transformation~(\ref{zetatrans}), we obtain the transformation rule
\be
\delta\gamma_{ij} = \xi^k\partial_k\gamma_{ij} + \gamma_{kj}\partial_i\xi^k +\gamma_{ik}\partial_j\xi^k -\frac{2}{3}\partial_k\xi^k\gamma_{ij}\,.
\label{delgam}
\ee
This is automatically traceless, using the fact that $\xi^k$ is a conformal Killing vector on $\mathbb{R}^3$,
\be
\delta^{ij} \delta\gamma_{ij} = 2\gamma_{ij} \partial^i\xi^j = \frac{2}{3}\partial_k\xi^k \gamma^i_{\ i}  = 0\,.
\ee
However, the transformation is not manifestly transverse:
\be
\partial^i \delta\gamma_{ij} = \gamma_{ij}\vec{\nabla}^2 \xi^i + \gamma_{ik}\partial^i\partial_j \xi^k - \frac{2}{3}\gamma_{ij} \partial^i\partial_k\xi^k\,.
\label{transverse}
\ee

For dilatations, the transverse condition $\partial^i\delta\gamma_{ij} = 0$ is indeed satisfied, since the gauge parameter is linear
in the coordinate, $\xi^i = \lambda x^i$. Thus dilatation is a symmetry even in the presence of tensors. Indeed, in this case~(\ref{delgam})
simplifies to
\be
\delta_\lambda \gamma_{ij} = \lambda x^k\partial_k\gamma_{ij}\,,
\label{deltagamma}
\ee
hence $\gamma_{ij}$ transforms as a scalar under dilatation. This implies the general transformation
\be
\delta_\lambda \left(e^ \gamma -\mathds{1} \right)_{ij} = \lambda
x^k\partial_k\left(e^ \gamma -\mathds{1} \right)_{ij}\,,
\label{deltagamma2}
\ee
which ensures that~(\ref{symchoicereqtensors}) extends to all orders in $\gamma$. In other words, dilatation is a symmetry
including tensor perturbations to all orders\footnote{This is consistent with Maldacena's derivation of a 3-point consistency relation
involving a long-wavelength scalar mode coupled to two short-wavelength tensor modes. As mentioned in the Introduction, such consistency relations
should be Ward identities for non-linearly realized dilatation
symmetry.},
just as it is a symmetry for $\zeta$ to all orders.

For SCTs, on the other hand, the gauge parameter $\xi^i = 2 b^kx_k x^i  - \vec{x}^2 b^i$ implies
\be
\partial^i \delta\gamma_{ij} = -6b^i \gamma_{ij}\,.
\ee
Special conformal transformations move us away from the transverse condition, and require a compensating gauge transformation. Hence they are not a good symmetry
in the presence of tensors, already at linear order in $\gamma$. They are an approximate symmetry of scalar perturbations, to the extent that tensor
modes can be neglected.

\subsection{Non-linearly realized tensor symmetries}
\label{NLtensors}

Tensor perturbations introduce their own set of non-linearly realized symmetries. We will find an infinite number of such symmetries, at least at the linear level in $\gamma$:
\be
h_{ij}=a^2(t)e^{2\zeta(\vec{x},t)} \(\delta_{ij} + \gamma_{ij}\)\,.
\ee
They will describe large gauge transformations that preserve the transverse, traceless conditions~(\ref{TT}). 

Consider a general linear diffeomorphism,
\be
\delta \gamma_{ij}=\partial_i\xi_j+\partial_j\xi_i\,.
\label{deltensor}
\ee
Because $\gamma_{ij}$ shifts non-linearly, to lowest order in $\gamma$ we focus on the non-linear part of the transformation. Meanwhile,
the curvature perturbation transforms linearly, $\delta\zeta = \xi^i\partial_i\zeta$. Taylor expanding around the origin, we can write
\be 
\xi_i=M_i+M_{i\ell_1}x^{\ell_1}+{1\over 2}M_{i\ell_1 \ell_2}x^{\ell_1}x^{\ell_2}+{1\over 3!}M_{i\ell_1 \ell_2\ell_3}x^{\ell_1}x^{\ell_2}x^{\ell_3}+\ldots\,,
\label{xiexpansion}
\ee
where the constant coefficients $M_{i\ell_1\cdots \ell_n}$ are symmetric in the last $n$ indices. The change in the metric is
\be
\delta \gamma_{ij} = M_{ij} + M_{ji} + \sum_{n = 2}^\infty \frac{1}{(n-1)!}\left(M_{i j \ell_2 \ldots \ell_n} + M_{j i \ell_2 \ldots \ell_n}\right) x^{\ell_2}\cdots x^{\ell_n}\,.
\label{deltensor2}
\ee
We have separated the $n=1$ term, for reasons that will be clear shortly. 

For this to preserve transversality and tracelessness of $\gamma$, we must have
\bea
\nonumber
M_{i i \ell_2 \ldots \ell_n} &=& 0\,;\\
M_{j i i \ell_3 \ldots \ell_n} &=& 0\,.
\label{MTTcond}
\eea
The latter of these conditions implies that $M_{i\ell_1\cdots \ell_n}$, with $n \geq 2$, is both symmetric and traceless in any two of its last $n$ indices. 
This so far leaves us with 
\be
3\times\left[ \left( \begin{array}{c}n+2 \\
n \end{array} \right) -  \left( \begin{array}{c}n \\
n-2 \end{array} \right) \right] = 3(2n + 1)
\ee
algebraically independent components at each order $n$. We have yet to impose the first of~(\ref{MTTcond}). This trace condition is manifestly symmetric in the last $n-1$ indices,
and hence imposes $2n-1$ constraints. The number of algebraically independent components is therefore $3(2n+1) - (2n-1) = 4(n+1)$, with $n\geq 2$. The case $n = 1$ is exceptional: it is clear from~(\ref{deltensor2}) that only the symmetric part of $M_{ij}$ contributes to the metric, leaving 5 independent components in this case. To summarize,
\be
[M_{i \ell_1 \ell_2 \ldots \ell_n}] 
=\left\{\begin{array}{cl}
5 \hspace{20pt}&\text{for}\hspace{10pt} n=1 \\ \\
4(n+1) \hspace{20pt}&\text{for}\hspace{10pt} n \geq 2\,.
\end{array}\right.
\label{Mcounting}
\ee 
We have this many symmetries at order $n$ of the expression~(\ref{xiexpansion}) --- an infinite number in total\footnote{Note that we have not imposed $M_{i\ell_1\ell_2...} = M_{\ell_1 i \ell_2...}$.
One might worry that in doing so, we have over counted, since there might be a certain subset of diffeomorphisms that give
an antisymmetric $\partial_j \xi_i$, {\it i.e.} they do not shift $\gamma_{ij}$ nonlinearly and should not be counted.
It can be shown that such diffeomorphisms do not exist, namely that an $n+1$-index object $M$ which is symmetric among its
last n indices, {\it and} is antisymmetric between its first two (or more
generally, antisymmetric between its first and any of the last $n$ indices),
must vanish. The proof goes by permuting indices
on $M_{i\ell_1 \ell_2 ...}$ to obtain $M_{\ell_2 \ell_1 i...}$ --- there are two
ways to do so but they yield opposite overall signs. This proof works
only for $n \ge 2$, which is why the $n=1$ case is special.
It is worth emphasizing that this proof does not mean
one should simply consider $M$ that is overall symmetric in all indices.
Consider a general $M$ that is symmetric among its last $n$ indices (but otherwise
with no further constraint on the first two indices), one can
write it as:
$M_{i\ell_1 \ell_2 ... \ell_n} = (M_{i\ell_1 \ell_2 ... \ell_n} + M_{\ell_1 i \ell_2 ... \ell_n} )/2
+ (M_{i\ell_1 \ell_2 ... \ell_n} - M_{\ell_1 i \ell_2 ... \ell_n})/2$.
One might be tempted to throw away the second term based on the
above argument, but if one defines $\tilde M_{i\ell_1 \ell_2 ... \ell_n}
\equiv  (M_{i\ell_1 \ell_2 ... \ell_n} - M_{\ell_1 i \ell_2 ... \ell_n})/2$, it
is clear in general $\tilde M$ is {\it not} symmetric among its last $n$ indices,
and so one cannot use the above proof to argue $\tilde M$ vanishes.}. 

As a check, we can similarly expand a general transverse, traceless metric $\gamma_{ij}$ as a power series around the origin. It is straightforward to show that
each order $n$ coefficient has $n(n+4)$ algebraically independent components. For $n = 1$ and $n=2$, this matches
the counting of gauge parameters in~(\ref{Mcounting}), which confirms that the metric and its first derivative ({\it i.e.}, Christoffel connection)
at a point can be completely gauged away. At $n =3$, corresponding to ${\cal O}(x^2)$ contributions in the metric, there are $21$ metric coefficients minus $16$ gauge parameters,
leaving us with $5$ physical components. This matches the number of curvature components at a point --- in 3 dimensions, curvature is fully
specified by the Ricci tensor, which in this case has 5 independent components since the Ricci scalar vanishes identically for a TT mode.

The lowest-order ($n=0$) transformation $\xi^i = M^i$ is just a translation, which trivially is a symmetry. 
The first non-trivial transformation  ($n=1$) is
\be
\xi^i = M^i_{\ j}x^{j}\,;\qquad  M^i_{\ i} = 0\, , \ M_{ij}=M_{ji}\, ,
\ee
{\it i.e.}, an anisotropic scaling of coordinates. This shifts the tensor perturbation by a constant matrix and corresponds to the usual constant adiabatic mode for the tensors~\cite{Weinberg:2008zzc}.
As shown in~\cite{Maldacena:2002vr}, the symmetry under anisotropic rescaling leads to a consistency relation involving the 3-point function of 
a long-wavelength tensor mode coupled to short-wavelength $\zeta$ modes. The infinite number of higher-order transformations in~(\ref{xiexpansion})
may be equally good symmetries. Large gauge transformations for tensors can readily be extended to physical mode solutions~\cite{Weinberg:2008zzc}. Indeed, for tensors 
there are no additional constraint equations to consider. It is therefore expected that~(\ref{xiexpansion}) should lead to further consistency
relations for correlation functions involving a tensor mode. 
 
\section{Discussion} 
\label{discussion}

To summarize, we have shown
that with a co-moving gauge choice of (\ref{comovgauge}),
there are certain residual symmetries that involve 
spatial transformations which do not vanish at infinity:
\begin{itemize}

\item An overall dilatation
$x^i \rightarrow  (1 + \lambda) x^i$, under which $\zeta$
transforms non-linearly (\ref{zetadil}) while the tensor transforms
linearly (\ref{deltagamma2}). This symmetry holds even for
large scalar and tensor metric perturbations, and has
been noted by previous authors.

\item A special conformal transformation
$x^i \rightarrow x^i + 2\vec x\cdot \vec b\, x^i -b^i \vec{x}^2$,
under which $\zeta$ transforms non-linearly (\ref{zetaSCT}).
This symmetry holds even for large $\zeta$, but requires
vanishing tensor perturbations.

\item An infinite set of symmetries (\ref{xiexpansion}) 
under which $\zeta$ transforms linearly, but the tensor does not.
These symmetries hold only for small ({\it i.e.}, linearized) tensor perturbations.

 \end{itemize}

Since these symmetries have a gauge origin (diffeomorphisms), 
a natural question is whether they have any physical significance.
We do not have a completely satisfactory answer to this question.
However, several lines of evidence suggest the answer is affirmative.

First, the dilatation symmetry has been used in a background
wave argument to derive the consistency relation (\ref{consis}) between
the three and two-point functions of $\zeta$ 
\cite{Maldacena:2002vr,Creminelli:2004yq,Cheung:2007sv}. Recently,
Creminelli, Nore\~na and Simonovi\'c ~\cite{Creminelli:2012ed} have derived
a consistency relation for the SCT symmetries. 

Second, as mentioned earlier, the symmetries of interest can be smoothly extended to physical perturbations at finite momenta, with suitable falloff conditions
at spatial infinity. For the conformal symmetries of the scalar sector, this was shown in Sec.~\ref{adiamodes}, while the corresponding argument for tensor symmetries
was mentioned in Sec.~\ref{NLtensors}.

A third evidence that suggests physical significance for our symmetries comes from considering the decoupling limit. The effective theory of inflation~\cite{Cheung:2007st} is a general framework for studying generic perturbations during inflation, and encompasses ordinary slow-roll inflation as a special case. In the effective theory, there is a field $\pi$ which is a St\"ukelberg field, introduced to restore time re-parametrization invariance into the formalism.  It is pure gauge, because it shifts under a time reparametrization. But there is a limit, the decoupling limit defined by $M_{\rm Pl}^2 \rightarrow \infty,~\dot H \rightarrow 0,~\dot HM_{\rm Pl}^2 \rightarrow {\rm const},$ in which the gauge transformation drops off of $\pi$, and $\pi$ becomes a physical field. The relation between $\zeta$ and $\pi$ to ${\cal O}(\pi^2)$ is $\zeta = -H\pi$.

In the decoupling limit, space-time becomes exactly de Sitter. In this case, there is a global $SO(4,1)$ symmetry for all the fields including $\pi$, corresponding to the isometries of the de Sitter on which they live.
Aside from spatial translations and rotations, these isometries include a space-time dilatation, $\tau \rightarrow (1+\lambda)\tau$, $x^i \rightarrow (1+\lambda)x^i$, where $\tau$ is conformal time, as well as 
$\tau \rightarrow (1+2\vec{b}\cdot\vec{x}) \tau$, $x^i \rightarrow x^i + 2\vec{x}\cdot\vec{b}x^i- b^i(-\tau^2 + \vec{x}^2)$, where the spatial part reduces to a SCT at late times ($\tau\rightarrow 0$), as it should. The St\"ukelberg field $\pi$ transforms non-linearly under these symmetries:
\bea
\nonumber
\delta_\lambda \pi &=& -\frac{\lambda}{H}(1+\dot\pi)+\lambda \vec x\cdot\vec\nabla\pi\,; \\
\delta_{b}\pi &=& -\frac{2}{H}\vec{b}\cdot\vec{x}(1 + \dot\pi) +\left(2\vec{b}\cdot\vec{x}x^i - \vec{x}^2b^i + \frac{b^i}{a^2H^2}\right)\partial_i\pi \,.
\eea
Translated to $\zeta$ language, these are precisely the spatial conformal symmetries we are considering in this paper.  In the decoupling limit, all the non-local operators in the effective theory disappear, and since the conformal transformations appear simply as isometries of the background, they should be physical symmetries under any reasonable definition.  Furthermore, if the process of taking the decoupling limit is itself smooth and physical, these symmetries should remain physical away from the decoupling limit.

In forthcoming work, we will explore the consequences of these symmetries on correlation functions.
It should be possible to understand the consistency relations as Ward identities for the broken scalar
and tensor symmetries, analogously to the low energy theorems of chiral perturbation theory. The infinite set of
tensor symmetries derived in Sec.~\ref{tensyms}, in particular, may give rise to novel consistency relations
involving tensors.

{\bf Acknowledgements:} This work has greatly benefited from many discussions and initial collaboration with Paolo Creminelli, Walter Goldberger, Alberto Nicolis, Jorge Nore\~na and Marko~Simonovi\'c.
We are grateful to Paolo Creminelli, Jorge Nore\~na and Marko~Simonovi\'c for sharing their preprint with us ahead of time, for feedback on an earlier version of this paper and in particular for pointing out a mistake in our earlier counting of tensor symmetries in Sec.~\ref{NLtensors}. We also thank Austin Joyce, Juan Maldacena,  Massimo Porrati, Leonardo Senatore, Sarah Shandera, Mark Trodden and Matias Zaldarriaga for helpful discussions. The work of K.H. and J.K. is supported in part by funds from the University of Pennsylvania, NASA ATP grants NNX11AI95G and NNX08AH27G, and the Alfred P. Sloan Foundation. The work of L.H. is supported in part by the DOE (DE-FG02-92-ER40699) and NASA (NNX10AN14G).

\end{document}